\documentstyle[aps,twocolumn,prb]{revtex}
\begin{document}
\draft

\title{Magnetization switching in nanoscale ferromagnetic grains: \\
       simulations with heterogeneous nucleation}

\author{M. Kolesik}
\address{
Supercomputer Computations Research Institute, 
Florida State University, Tallahassee, Florida 32306-4052
        }

\author{Howard L.\ Richards}
\address{
Department of Physics,  University of Tokyo, Hongo, Bunkyo-ku, 
Tokyo 113, Japan
        }

\author{M. A. Novotny}
\address{  
Supercomputer Computations Research Institute, 
Florida State University, Tallahassee, Florida 32306-4052 \\
and
Department of Electrical Engineering,
2525 Pottsdamer Street,   
Florida A\&M University--Florida State University,
Tallahassee, Florida 32310-6046
        }

\author{Per Arne Rikvold}
\address{  
Center for Materials Research and Technology,
Department of Physics, \\
and 
Supercomputer Computations Research Institute, 
Florida State University, Tallahassee, Florida 32306-3016 \\
and
Department of Fundamental Sciences, Faculty of Integrated Human Studies, 
Kyoto University, Kyoto 606-01, Japan
        }

\author{Per-Anker Lindg\aa rd}
\address{        
Department of Solid State Physics, Ris{\o}~National Laboratory,
DK-4000 Roskilde, Denmark
        }

\date{\today}
\maketitle
\begin{abstract}
We present results obtained with various types of 
heterogeneous nucleation in a kinetic Ising model of magnetization 
switching in single-domain ferromagnetic nanoparticles. 
We investigate the effect of the presence of the system boundary
and make comparison with simulations on periodic lattices.
We also study systems with bulk disorder and compare how two different
types of disorder influence the switching behavior.
\end{abstract}


This work is devoted to computer simulation of magnetization switching 
in single-domain ferromagnetic nanoparticles.
For highly anisotropic systems, simple kinetic Ising models can 
qualitatively explain many experimental observations.\cite{MMM} 
However, to obtain more realistic results additional physical 
effects must be included in the model.  
An important aspect of real samples is heterogeneous
droplet nucleation due to the presence of the system boundary and
due to bulk disorder.

The model used in our study is a square-lattice nearest-neighbor kinetic Ising
ferromagnet with random updates using either
Glauber or Metropolis single-spin-flip Monte Carlo dynamics.
We first investigate systems with boundaries (but without disorder),
then we consider disordered systems with periodic boundary conditions.
We concentrate on two quantities: the lifetime and the switching field.

To determine the metastable lifetime, we start the simulations 
with all spins $+1$ and impose a negative magnetic field. 
Then we measure (in Monte Carlo Steps per Spin, MCSS) 
the mean time  the sample needs to reach  zero magnetization.
Having measured this ``lifetime'' $\tau$ for several values of 
the magnetic field,
one can extract from the data the switching field $H_{\rm sw}$, 
which is defined as the field that produces a given lifetime. 
This quantity is often measured in experiments.\cite{Chang,Lederman,New}

We performed Monte Carlo simulation studies of the switching dynamics
of a kinetic Ising model defined on circular lattices (which we define as
subsets of a square lattice, contained within a circle of given diameter $L$).
The Hamiltonian is 
\begin{equation}
{\cal H} = -J\sum_{<ij>} \sigma_i \sigma_j 
    -H \sum_i \sigma_i 
    -H_\Sigma \sum_{i \in \Omega} \sigma_i \ ,
\end{equation}
where the first two terms represent the standard spin-spin interaction with
positive $J$ and the coupling to the external field $H$, respectively. The last term,
in which the summation runs only over the sites on the boundary of the
lattice $\Omega$, is included to model the effect of the system boundary
on nucleating droplets of the stable phase. For $H_\Sigma =0$ one has
a free boundary, whereas positive (negative) $H_\Sigma$ mimic a boundary which
effectively repels (attracts) the droplets.

\begin{figure}
\vspace*{2.7in} 
\includegraphics{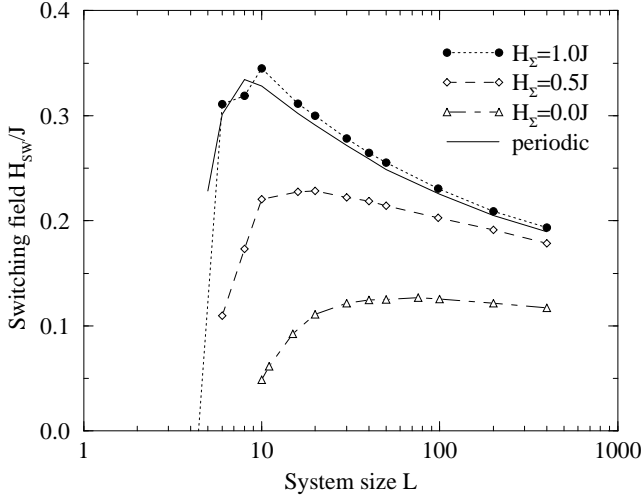} 
\caption{ 
The switching field $H_{\rm sw}$ vs.\ $L$ for different values of the boundary field
$H_{\Sigma}$ for circular systems at $T=1.3J \approx 0.57 T_c$.
The waiting time is $\tau=30000$ MCSS, the dynamics is Metropolis. 
Data for periodic systems (full line) are shown for comparison.
Note that even lattices with $L=400$ are still in the single-droplet region. 
The decrease of the switching field would continue 
if we went to even larger systems. Eventually, for
very large systems, it would converge to the same value
in the multidroplet region, independent of $H_\Sigma$. 
}
\label{fig:cirb}
\end{figure}

Figure~\ref{fig:cirb} shows the switching fields for a fixed waiting
time $\tau$ versus the diameter $L$ of the circular lattice.
For the smallest systems, $H_{\rm sw}$ is zero; 
for slightly larger systems, $H_{\rm sw}$ increases sharply with $L$.  
This is the coexistence region, in which
the stable and metastable phases practically coexist.\cite{MMM,Rikvold}
For larger systems, the behavior depends strongly on
the affinity of the boundary for droplets of the equilibrium phase, 
which here is modeled by different values of $H_\Sigma$; 
and also on the waiting time $\tau$. In general, the increase
in the coexistence region is followed by a decrease of the switching field
in the single-droplet region (where switching is triggered by a single
critical droplet\cite{MMM,Rikvold}). 
This results in a maximum located at the crossover between the 
coexistence and single-droplet regions.\cite{MMM}
Similar switching-field peaks are observed
in certain experiments on nanoscale ferromagnets.\cite{Chang} 
For larger systems, more than one 
droplet nucleates during the switching process, and the system is in 
the multidroplet region.\cite{MMM,Rikvold} There, the switching field 
becomes asymptotically independent of the system size.

Just as in periodic systems,\cite{MMM} this behavior can be clearly observed 
in samples in which the boundaries 
repel droplets of the stable phase (positive $H_\Sigma$). 
For neutral boundaries ($H_\Sigma=0$) the maximum in the switching field 
can be much less pronounced or it can disappear completely, 
depending on the waiting time. 
For example, with $\tau = 1000$, the maximum of $H_{\rm sw}$ completely disappears
for $H_\Sigma = 0$. In general, to observe the
maximum, one needs a longer waiting time. Note that a long 
waiting time probably corresponds better to most real experimental situations.
For theoretical considerations concerning magnetization switching
in kinetic Ising systems without disorder, 
see Refs.\onlinecite{MMM} and \onlinecite{Richards}.

To study effects of disorder,
we simulated the above Ising model on periodic
square lattices using Glauber dynamics. The first type of disorder we investigate
is produced by defects generated by randomly deleting bonds 
of the lattice with concentration $c$. 

In order to understand how the disorder influences the switching,
we measured various properties of the nucleating
droplets of the stable phase during the simulation. 
All measured quantities were taken as
mean values over all events in which a droplet of a given size appeared
in the system. It turns out that with this type of
disorder, the switching can be well described in terms of  ``average''
nucleating droplets. 

\begin{figure}
\vspace*{2.5in}
\includegraphics{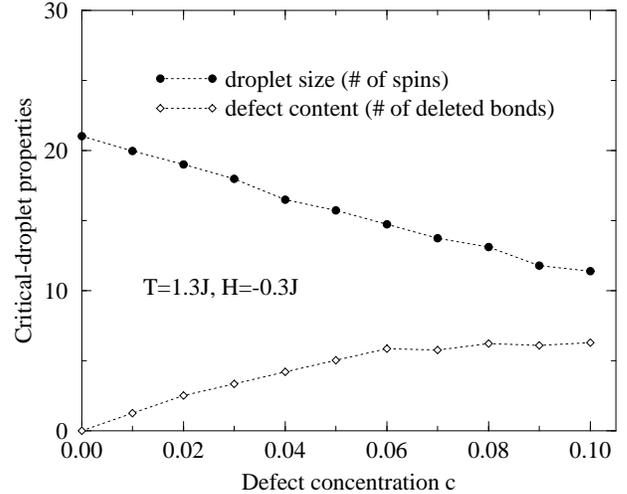} 
\caption{ 
Size and defect content of a critical droplet 
as  functions of the defect concentration.
}
\label{fig:cdpr}
\end{figure}

Figure~\ref{fig:cdpr} shows the critical size of a droplet versus
the defect concentration $c$. Droplets larger than the critical size
are more likely to grow than to shrink, whereas smaller droplets do the
opposite. As expected, the disorder leads to smaller critical droplets, 
which in turn results in a dramatic decrease of the metastable lifetime.

\begin{figure}
\vspace*{2.6in}
\includegraphics{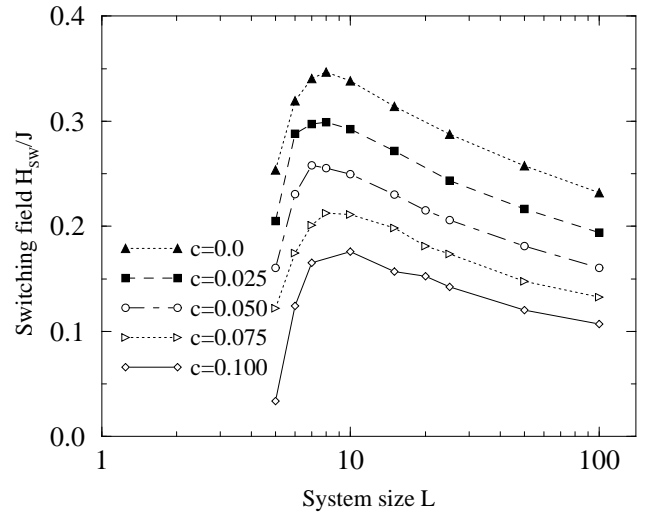} 
\caption{ 
Switching field as a function of system size for different 
defect concentrations, waiting time $\tau = 30000$ MCSS
and Glauber dynamics at $T=1.3J$. 
}
\label{fig:hswi}
\end{figure}

Another interesting property of the typical critical droplet is its 
defect content: the number of deleted bonds associated with
the droplet. It can be deduced from Fig.~\ref{fig:cdpr}, as well as
from direct measurements, that the effective concentration of defects
within the droplets is considerably larger than the global concentration. 
This is caused by the fact that the droplets are preferentially 
nucleated in the vicinity of the defects.

In Fig.~\ref{fig:hswi} we show the effect of the disorder on the switching
field. Whereas the switching field is decreased by the disorder, its shape
as a function of the system size remains approximately independent of $c$.

Thus, in the intermediate-temperature region studied here, 
the switching dynamics of
the kinetic Ising model on a bond-diluted lattice is essentially the
same as in  systems without disorder, although the metastable lifetimes
and switching fields are considerably reduced by the disorder. The effect of 
the disorder can be approximately described in terms of an effective
medium which affects the growing droplets of the stable phase. Details will
be given in a future paper.\cite{Richards}

\begin{figure}
\vspace*{2.65in}
\includegraphics{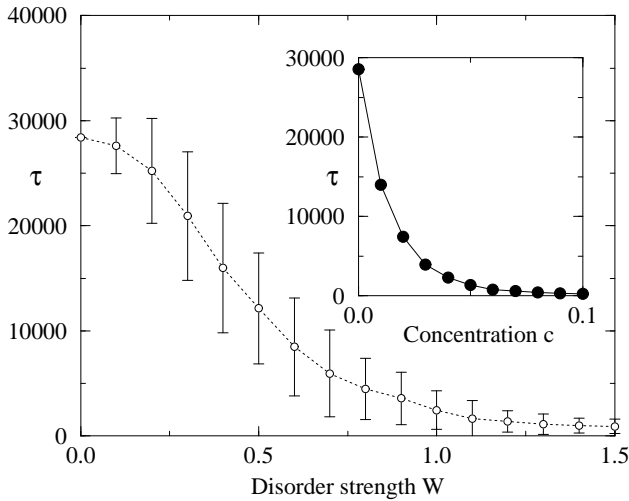} 
\caption{ 
The metastable lifetime $\tau$ as a function of disorder.
The error bars show the width of the distributions of lifetimes for 
different realizations of the disorder.  
For the diluted-bond type of disorder (insert),
this broadening is smaller than the symbol size.
In both cases, $T=1.3J$, $H=-0.3J$ and $L=20$.
\label{fig:tIItau}
}
\end{figure}

Another type of disordered kinetic Ising model is
defined by the Hamiltonian
$
{\cal H} = -J\sum_{<ij>} a_i a_j \sigma_i  \sigma_j - H \sum_i a_i \sigma_i 
$, 
where the amplitudes $a_i$ represent random local magnetic moments.
Here, we take them uncorrelated with a ``box'' distribution of  width $W$, 
centered around unity.
The switching dynamics with this type of disorder differs from
the one discussed above in several respects. First, the approach 
in terms of ``average'' nucleating droplets is not useful 
because the interesting quantities have broad distributions.
Second, there is a lack of self-averaging, even in the
intermediate-temperature region studied here. This is demonstrated in 
Fig.~\ref{fig:tIItau}, where we show how the mean lifetime is reduced
by the disorder. At the same time, the relative width of the probability distribution
of the mean individual-sample lifetimes increases. Whereas switching 
of a sample with a particular realization of the disorder is more 
deterministic than in
a pure system, an ensemble of disordered systems exhibits a wide
distribution of lifetimes and, consequently, of switching fields.
The mechanism leading to this behavior is that the size of a critical
droplet is different in different parts of the disordered lattice.

In summary, we have studied magnetization switching in systems in which
heterogeneous nucleation of the equilibrium phase may occur on the system
boundary or be associated with bulk impurities.

The presence of the system boundary strongly affects the magnetization switching
in small systems since the critical fluctuations of the stable phase tend to
nucleate in its vicinity. This considerably decreases the metastable lifetime 
as well as the switching field, compared to periodic systems. 
However, the basic features, such as the crossover from the coexistence region
to a single-droplet region, and then to a multidroplet
region with increasing $L$, can still be observed in systems with a boundary. 
The switching field as a function of the system size becomes less peaked, 
but a maximum, located near the crossover between the coexistence and the 
single-droplet regions, can still be observed if the waiting time is 
sufficiently long. 

We have studied two different types of disordered systems. 
Diluted-bond disorder
can be a toy model for samples with impurities, whereas the model with
random magnetic moments could represent films with fluctuating thickness.
Our results show that whereas these types of bulk disorder result in shorter metastable
lifetimes in general, the details of the magnetization switching can differ 
considerably for different types of disorder.

Research supported by FSU-MARTECH, FSU-SCRI 
(DOE Contract No. DE-FC05-85ER25000),
NSF Grants No. DMR-9315969, DMR-9520325, and INT-9512679, 
the Inoue Foundation, and NERSC supercomputer time.


%
%

\end{document}